\documentclass[twocolumn,secnumarabic,amssymb, nobibnotes, aps, prl]{revtex4-1}

\usepackage{graphicx}

\begin{document}

\title{Ultra-High Reynolds Number Is Not Necessary for Comprehensive Log Scaling in a Turbulent Boundary Layer}

\author{Shivsai Ajit Dixit$^{1}$}%
\author{O N Ramesh$^{2}$}
\affiliation{$^{1}$Indian Institute of Tropical Meteorology, Pashan, Pune 411008, India\\$^{2}$Indian Institute of Science, Bangalore, Karnataka 560012, India}
\date{}%

\begin{abstract}

\noindent Experiments in an extraordinary turbulent boundary layer called the sink flow, displaying a perfect streamwise invariance, show a wide extent of logarithmic scaling for moments of streamwise velocity up to order 12, even at moderate Reynolds numbers. This is in striking contrast to canonical constant-pressure turbulent boundary layers that show such comprehensive log scaling only at ultra-high Reynolds numbers. This demonstrates that ultra-high Reynolds number is not necessary for comprehensive log scaling to occur; while specific details of the sink flow are special, the relevance to general turbulent boundary layers is that the sink flow underscores the importance of the streamwise invariance condition that needs to be met in a general flow for obtaining log scaling. Indeed, a simple heuristic theory shows that, for log scaling in the inertial sublayer, the invariance of dimensionless mean velocity and higher-order moments along a mean streamline is a necessary and sufficient condition. Ultra-high Reynolds number primarily appears to make these log scalings universal. 
\end{abstract}

\maketitle

A celebrated theoretical result in the literature on wall-bounded turbulent flows is the logarithmic scaling of mean velocity with distance from the wall \cite{Millikan}. While this has been the object of much scrutiny in the literature for the past few decades with competing claims from power-law scaling \cite{Barenblatt1993}, the log scaling has remained robust and has stood the test of time. Data from high-Reynolds-number (high-Re) experiments have started becoming available only in the recent times. These measurements show that the log region at ultra-high Reynolds numbers is more comprehensive - not only the mean velocity but also the streamwise intensity varies logarithmically there \cite{Hultmarketal2012,Marusicetal2013}. Higher-order even moments of streamwise velocity fluctuation also show log scaling over this region \cite{MeneveauMarusic2013,Vallikivietal2015}.

The conventional wisdom is that a turbulent boundary layer (TBL) flow is a two-lengthscale problem - there is an inner layer where the lengthscale is viscous ($\nu/u_{\tau}$, where $\nu$ is the kinematic viscosity of fluid and $u_{\tau}$ is the so-called friction velocity, to be defined shortly) and a distinct outer layer where the dynamics is essentially inviscid and boundary layer thickness $\delta$ is the lengthscale there. The velocity scale in both the layers is $u_{\tau} = \tau_{w}/\rho$ ($\tau_{w}$ is the wall-shear stress and $\rho$ is the fluid density), whereas the independent variables are $y_{+} = yu_{\tau}/\nu$ and  $\eta = y/\delta$ respectively; $y$ is the distance from the wall. The Reynolds number $\delta_{+} = \delta u_{\tau}/\nu = y_{+}/\eta$ measures the disparity between the largest and smallest eddy lengthscales. If $\delta_{+} \rightarrow \infty$, one can expect an inertial sublayer (ISL) where both inner and outer scaling descriptions are simultaneously valid \cite{RN1990}. This corresponds to the simultaneous limits $y_{+} \rightarrow \infty$ and $\eta \rightarrow 0$, leading to $y$ as the only lengthscale in the ISL \cite{TennekesLumley}. As a dimensional necessity, $\partial\left<u\right>/\partial y \sim u_{\tau}/y$ in the ISL which yields the log scaling of mean velocity $\left<u\right>$. Due to its asymptotic nature, the log scaling is seen as an infinite-Re expectation; there have been attempts to make finite-Re corrections to the traditional asymptotic theory (see \cite{Afzal1976}, for example) but we will not pursue them here. The following questions, however, remain unaddressed: (1) How do we understand log scaling of mean velocity in moderate-Re flows? (2) What is the physical content of the so-called law-of-the-wall for mean velocity $\left<u_{+}\right> = \left<u/u_{\tau}\right> = f\left(y_{+}\right)$, a generalized functional form of the log scaling? (3) How do we understand log scaling for higher-order moments of velocity? 

\begin{figure}
\includegraphics[scale=0.35]{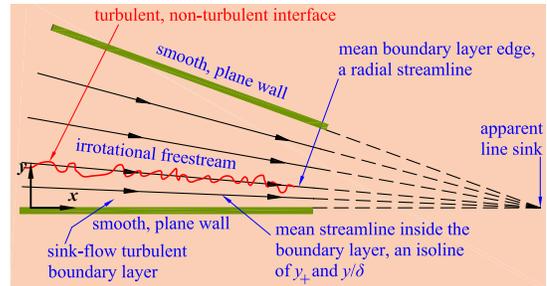}
\caption{Schematic of a sink-flow turbulent boundary layer.\label{fig:1}}
\end{figure}

In this Letter, we seek to address these questions in the context of streamwise velocity component and present a unified picture of log scaling in TBL flows. We show that a necessary and sufficient condition for the occurrence of comprehensive log scaling in TBL flows is the invariance of dimensionless mean velocity and higher-order moments along a mean streamline (streamwise invariance) in the ISL. This condition is consistent with the law-of-the-wall kind of scaling for mean velocity and higher-order moments as will be shown later. For mean velocity, the physical implication is that the iso-$y_{+}$ lines (or iso-$\eta$) lines in the ISL coincide with the mean streamlines i.e. there is no mean exchange (`entrainment') of fluid between the outer wake region and the inner region (including ISL). It is shown that for canonical constant-pressure TBLs, the above condition can be attained in the ISL only at ultra-high Reynolds numbers ($\delta_{+}$ of order $10^{4}$ or larger). However it tuns out that in order to obtain an extended and comprehensive log scaling, even a moderate Reynolds number, that ensures an ISL, is enough, provided the flow geometry is such that the above condition is satisfied irrespective of Reynolds number. We demonstrate our contention using experimental data from a seemingly special configuration - the sink-flow TBL (Fig.~\ref{fig:1}) - where flow geometry ensures streamwise invariance and the flow displays dramatic log scalings for all moments of streamwise velocity (up to order 12) at Reynolds numbers $\delta_{+}$ that would have been considered only moderate (order $10^{3}$) by traditional wisdom. Moreover, the region of comprehensive log scaling extends over a considerable wall-normal extent that is unencumbered by the so-called edge intermittency effects. For canonical constant-pressure TBLs, simultaneous log regions of mean velocity and streamwise intensity are observed to emerge only at ultra-high Reynolds numbers ($\delta_{+}$ of order $10^{4}$) \cite{Marusicetal2013,MeneveauMarusic2013,Vallikivietal2015} - an order of magnitude larger than the sink-flow Reynolds number. This region corresponds to log scaling of even moments as well \cite{MeneveauMarusic2013,Vallikivietal2015}. It is to be noted that moderate-Re constant-pressure TBLs show log scaling for mean velocity but not for intensity (and other higher-order moments).

The sink-flow TBL (Fig.~\ref{fig:1}) is characterized by an exactly self-similar downstream development \cite{Townsend1976} such that all locally-defined Reynolds numbers (e.g. $\delta_{+}$) essentially remain invariant in the streamwise direction \cite{DixitRamesh2008}. It is emphasized that the sink flow is only seemingly special and conclusions from this flow have general applicability to other TBL flows; the special geometry of sink flow should be seen merely as a device to ensure the streamwise invariance condition (not only in the ISL but all across the TBL). Indeed, due to the remarkable streamwise invariance it displays, the sink flow could serve as a convenient reference that can throw light on the conditions required for log scaling in a general TBL flow, without getting cluttered by Reynolds number issues (much like its laminar counterpart - Falkner-Skan flows). From the perspective of the so-called attached eddy hypothesis, a sink-flow TBL contains only wall eddies (type A) whereas a canonical constant-pressure TBL contains extra wake eddies (type B) also \cite{Perryetal2002}. It is to be noted that wake eddies do not contribute to (but actually obfuscate) the log scaling; wall eddies alone suffice to yield the log scaling of mean velocity and intensity (see \cite{PerryMarusic1995} and figure 5\textit{a} of \cite{MarusicPerry1995}). In that sense, the structure of the sink flow is minimalistic and serves as a substratum for more general TBL flows. 

Sink-flow experiments reported here were performed in an open-return, suction-type wind tunnel. Test-section of the tunnel was 1 foot $\times$ 1 foot in cross-section with a working length of 3 meters. Sink-flow geometry was created by appropriately contouring the ceiling of the test-section. Boundary layer developing on an aluminium test-surface was measured by Pitot-tubes (for mean velocity $\left<u\right>$) and hotwire anemometry (for velocity fluctuation $u'$); $u = \left<u\right> + u'$ where $u$ is the instantaneous velocity. Details of the setup and instrumentation can be found in \cite{DixitRamesh2008,DixitRamesh2010,DixitRamesh2013}. Table~\ref{table1} shows summary of experimental parameters for the present sink-flow TBL.

\begin{table}
\caption{\label{table1} Parameters for the present sink-flow turbulent boundary layer. $u_{\infty}$, freestream velocity $(\textrm{in ms}^{-1})$; $u_{\tau}$, friction velocity $(\textrm{in ms}^{-1})$; $\delta_{+} = \delta u_{\tau}/\nu$, Reynolds number where $\delta$ is boundary layer thickness corresponding to $\left<u\right> = 0.99u_{\infty}$; $K = \left(\nu/u_{\infty}^{2}\right)\textrm{d}u_{\infty}/\textrm{d}x$, acceleration parameter; $l/d$, length-to-diameter ratio of hotwire sensor; $l_{+} = lu_{\tau}/\nu$, sensor length in viscous units; $Tu_{\infty}/\delta$, dimensionless duration of data stretch.}
\begin{ruledtabular}
\begin{tabular}{ccccccc}
$u_{\infty}$      & $u_{\tau}$   &  $\delta_{+}$ & $K\times 10^{7}$ & $l/d$ & $l_{+}$ & $Tu_{\infty}/\delta$ \\ \hline     
$20.43$ & $0.9122$ & $828$ & $7.71$ & $160$ & $49$ & $92228$\\
\end{tabular}
\end{ruledtabular}
\end{table}        

\begin{figure}
\includegraphics[scale=0.28]{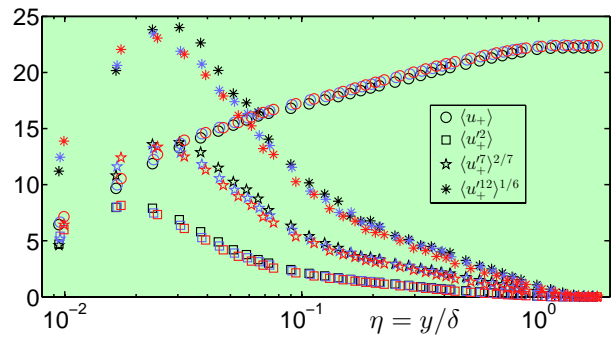}
\caption{Streamwise scaling of profiles of mean velocity $\left<u_{+}\right>$ and higher moments of $u'$ (order 2, 7 and 12 are only shown to avoid clutter) in the present sink flow. Symbols represent different moments and colors represent three streamwise measurement stations with distance between adjacent stations being four times the typical boundary layer thickness \cite{DixitRamesh2008}. \label{fig:2}}
\end{figure}

\begin{figure*}
\includegraphics[scale=0.33]{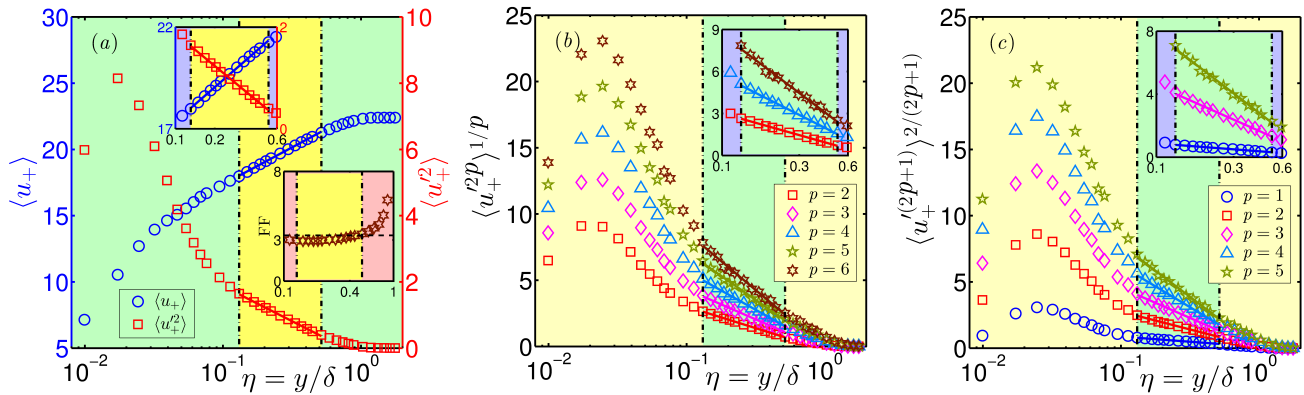}
\caption{Profiles of (\textit{a}) mean velocity $\left<u_{+}\right>$ and intensity $\left<u'^{2}_{+}\right>$, (\textit{b}) even moments $\left<u'^{2p}_{+}\right>^{1/p}$ (order 4 to 12, $p = 2$ to $6$) and (\textit{c}) odd moments $\left<u'^{2p+1}_{+}\right>^{2/2p+1}$ (order 3 to 11, $p = 1$ to $5$) in the present sink flow. Full lines show least-squares logarithmic fits to the data. Extent of the log region is marked by vertical dashed-dotted lines. Each upper inset shows enlarged view of the log region in representative profiles. Right inset in (\textit{a}) shows wall-normal variation of flatness factor $\textrm{FF} = \left<u'^{4}\right>/\left<u'^{2}\right>^{2}$ of $u'$; dashed line marks the FF value corresponding to $\gamma = 3/\textrm{FF} = 0.9$, $\gamma$ being the edge-intermittency factor \cite{McComb1991}. \label{fig:3abc}}
\end{figure*}

Figure~\ref{fig:2} demonstrates scaling of profiles of mean velocity $\left<u_{+}\right>$ and representative higher-order moments of $u'_{+}$ measured at three different streamwise stations; subscript $+$ refers to normalization using $u_{\tau}$. Remarkable collapse in both, outer and inner coordinates is observed (inner not shown). Regions of log scaling are also identifiable for each moment. Notice that the scaling is with respect to the streamwise development in a given sink flow where the Reynolds number remains invariant downstream. This is not to be confused with Reynolds number scaling in constant-pressure TBLs where Reynolds number increases downstream so that streamwise scaling and Reynolds number scaling are equivalent. Also, note that appropriate powers, such as $\left<u'^{7}_{+}\right>^{2/7}$ etc., of higher moments have been taken to express them such that they are of the same order as intensity (see also \cite{MeneveauMarusic2013}). Convergence (to within $3\%$) of higher-order moments has been verified (not shown) by computing $\left<u'^{m}_{+}\right>$ $\left(m = 2,3,\ldots,12\right)$ as area under the curve of $u'^{m}_{+}P\left(u'_{+}\right)$ where $P\left(u'_{+}\right)$ is the probability density function (pdf) of $u'_{+}$ \cite{MeneveauMarusic2013}.

We now present strong evidence for comprehensive log scaling in the present sink-flow TBL using data from a representative streamwise station. Figure~\ref{fig:3abc} shows profiles of mean velocity, streamwise intensity and even as well as odd moments of higher orders (3 to 12). An impressive coexistence of log variations of the form

\vspace{-10pt}
\begin{eqnarray}
\left< u_{+}\right> &=&\kappa^{-1}\ln \left(\eta\right) + B \label{eqn:meanvelloglawouter},\\
\left< u'^{2p}_{+}\right>^{1/p} &=& -A_{p}\ln \left(\eta\right) + C_{p}\label{eqn:evenmomentsloglawouter},\\
\left< u'^{2p+1}_{+}\right>^{2/2p+1} &=& -A^{*}_{p}\ln \left(\eta\right) + C^{*}_{p}\label{eqn:oddmomentsloglawouter},
\end{eqnarray}
\vspace{-10pt}

\noindent is immediately evident indicating comprehensive log scaling. Here, $p = 1,2,\ldots$ in Eqs.(\ref{eqn:evenmomentsloglawouter}) and (\ref{eqn:oddmomentsloglawouter}); $p = 1$ in Eq.(\ref{eqn:evenmomentsloglawouter}) yields the equation $\left<u'^{2}_{+}\right> = -A_{1}\ln \left(\eta\right) + C_{1}$ for streamwise intensity. Wall-normal distance $y$ has been normalized by the boundary layer thickness ($\eta = y/\delta$) although use of inner scaling is more common in the literature. It is important to note that the log fits in Fig.~\ref{fig:3abc} have been performed only over the region that is unaffected by edge intermittency and spatial resolution issues. The lower limit of this region is taken to be $y_{+} \approx 2.6\sqrt{\delta_{+}}$ (see \cite{MeneveauMarusic2013}); spatial resolution errors \cite{Smitsetal2011} are insignificant beyond this location. The upper limit is taken to be the wall-normal location beyond which the edge-intermittency factor ($\gamma$) falls below 0.9, indicating that the flow below this location is turbulent for more than $90\%$ of the duration; $\gamma = 3/\textrm{FF}$ \cite{McComb1991} where FF is the flatness factor of $u'$ (see right inset of Fig.~\ref{fig:3abc}\textit{a}). Rapid increase in FF value beyond this upper limit is consistent with the effects of edge intermittency \cite{DengelFernholz1990,FernholzFinley}. In effect, the width of the comprehensive log region in Fig.~\ref{fig:3abc} is $0.13\leq y/\delta \leq 0.52$. Recently, Huisman \emph{et al.} \cite{Huismanetal2013} have reported mean velocity and intensity log scalings in boundary layers of a turbulent Taylor-Couette flow; these scalings display striking coexistence over a range of wall-normal locations. Some cases from this study have Reynolds numbers that are remarkably low ($\textrm{Ta} = 1.5\times10^{10}$ case - $\textrm{Ta}$ is the Taylor number - has $\delta_{+} \approx 2\times10^{3}$; see Fig.~3\textit{a} of \cite{Huismanetal2013}), similar to the present study, and in contradistinction to canonical TBLs \cite{Marusicetal2013,Vallikivietal2015}. 

Coefficients $\kappa^{-1}$, $A_{p}$ and $A^{*}_{p}$ in Eqs.~(\ref{eqn:meanvelloglawouter}) to (\ref{eqn:oddmomentsloglawouter}) are expected to depend on the driving pressure gradient. Furthermore we expect these coefficients to approach universal values as $\delta_{+} \rightarrow \infty$. In sink flows, it is already known that $\kappa$ depends on pressure gradient and asymptotes to a value of about $0.4$ as the pressure gradient tends to zero with the corresponding increase in Reynolds number \cite{DixitRamesh2008}. For log scaling, such variations of coefficients with the driving parameters have been reported in other experiments as well \cite{Huismanetal2013}.

In order to heuristically understand these results, consider

\vspace{-10pt}
\begin{equation}
\textrm{ADV}\left(\left< b_{+}^{m}\right>^{n}\right)\triangleq\left< u\right>\frac{\partial \left< b_{+}^{m}\right>^{n}}{\partial x}+\left< v\right>\frac{\partial \left< b_{+}^{m}\right>^{n}}{\partial y}\label{eqn:meanadvdef},
\end{equation}
\vspace{-10pt}

\noindent which is advection of a generalized moment $\left<b_{+}^{m}\right>^{n}$ of streamwise velocity in a steady and two-dimensional flow where $\left<u\right>$ and $\left<v\right>$ are mean velocities in streamwise ($x$) and wall-normal ($y$) directions respectively. Physically, this represents spatial rate of change of $\left<b_{+}^{m}\right>^{n}$ along a mean-flow streamline and appears in the left side of the dynamical equation for $\left<b_{+}^{m}\right>^{n}$. Note that $\left<b_{+}^{m}\right>^{n} = \left<u_{+}\right>$ or $\left<u'^{2p}_{+}\right>^{1/p}$ or $\left<u'^{2p+1}_{+}\right>^{2/2p+1}$. For a sink-flow TBL, the scaling observed in Fig.~\ref{fig:2} implies that $\left<b_{+}^{m}\right>^{n}$ does not vary for a fixed value of $\eta$ (or $y_{+}$). Also, the radial pattern of mean streamlines (Fig.~\ref{fig:1}) leads to the invariance of $\eta$ and $y_{+}$ along a streamline \cite{Townsend1976,DixitRamesh2008,DixitRamesh2010}. Thus $\left<b_{+}^{m}\right>^{n}$ remains invariant along a mean streamline i.e. $\textrm{ADV}\left(\left<b_{+}^{m}\right>^{n}\right) = 0$. As will be shown shortly, this readily leads to comprehensive log scaling in sink flows. 

Is it possible that the streamwise invariance embodied in $\textrm{ADV}\left(\left<b_{+}^{m}\right>^{n}\right) = 0$ condition is responsible for log scaling, in general, and not just limited to sink flows? Towards this, we propose $\textrm{ADV}\left(\left<b_{+}^{m}\right>^{n}\right) = 0$ as an \emph{ansatz} for log scaling in general TBLs. With this, Eq.~(\ref{eqn:meanadvdef}) yields

\vspace{-10pt}
\begin{equation}
\frac{\partial \left< b_{+}^{m}\right>^{n}}{\partial y}=-\frac{\left< u\right>}{\left< v\right>}\frac{\partial \left< b_{+}^{m}\right>^{n}}{\partial x}\label{eqn:delbydely}.
\end{equation}
\vspace{-10pt}

\noindent Further, one may substitute the law-of-the-wall scaling $\left<u_{+}\right> = f\left(y_{+}\right)$ for $\left<u\right>$ in the mean mass conservation equation $\partial\left<u\right>/\partial x + \partial\left<v\right>/\partial y = 0$ to obtain $\left<u\right>/\left<v\right> = -\left[u_{\tau}/\left(\textrm{d}u_{\tau}/\textrm{d}x\right)\right]/y = -x_{s}/y_{s}$. Here $x_{s} = u_{\tau}/\left(\textrm{d}u_{\tau}/\textrm{d}x\right)$ and $y_{s} = y$ are the streamwise and wall-normal lengthscales respectively and $y_{s} = y$ is predicated on the existence of an ISL. In addition, the inner-scaled asymptotic expansion for $\left<b_{+}^{m}\right>^{n}$ may be written as $\left<b_{+}^{m}\right>^{n} = h_{0} + \epsilon_{1} h_{1} + \epsilon_{2} h_{2} + \cdots$ where $\epsilon_{i}=\epsilon_{i}\left(\delta_{+}\right)$ are gauge functions and $h_{0},h_{1}\left(y_{+}\right),h_{2}\left(y_{+}\right),\ldots$ are coefficients, $h_{0}$ being a constant of $\mathcal{O}(1)$ (convincingly demonstrated in \cite{MonkewitzNagib2015} for intensity). To the lowest order in $\epsilon$, $\partial\left<b_{+}^{m}\right>^{n}/\partial x$ then scales as $\mathcal{O}(1)/x_{s}$ and Eq.~(\ref{eqn:delbydely}) yields

\vspace{-10pt}
\begin{equation}
\frac{\partial \left< b_{+}^{m}\right>^{n}}{\partial y}\sim \frac{x_{s}}{y_{s}}\frac{\mathcal{O}\left(1\right)}{x_{s}}\sim\frac{\mathcal{O}\left(1\right)}{y}\label{eqn:delbydelyscaled}.
\end{equation}
\vspace{-10pt}
 
\noindent Integration of Eq.(\ref{eqn:delbydelyscaled}) yields $\left< b_{+}^{m}\right>^{n} \sim \ln(y)$. Expressed in outer coordinates, this reads  

\vspace{-10pt}
\begin{equation}
\left< b_{+}^{m}\right>^{n}=A_{m,n}\ln \left(\eta\right) + C_{m,n}\label{eqn:genloglawouter}.
\end{equation}
\vspace{-10pt}

\noindent Here $C_{m,n}$ and $A_{m,n}$ are coefficients; $A_{m,n}$ can be identified to be $\kappa^{-1}$ or $-A_{p}$ or $-A^{*}_{p}$ (Eqs.~\ref{eqn:meanvelloglawouter}, \ref{eqn:evenmomentsloglawouter}, \ref{eqn:oddmomentsloglawouter}). This shows that $\textrm{ADV}\left(\left<b_{+}^{m}\right>^{n}\right) = 0$ is in fact a sufficient condition for the occurrence of log scaling of mean velocity and higher moments in the ISL of general TBL flows.

In order to show that $\textrm{ADV}\left(\left<b_{+}^{m}\right>^{n}\right) = 0$ is a necessary condition for log scaling of $\left<b_{+}^{m}\right>^{n}$ in the ISL of a general TBL flow, one needs to show that if $\textrm{ADV}\left(\left<b_{+}^{m}\right>^{n}\right) \neq 0$, log scaling of $\left<b_{+}^{m}\right>^{n}$ cannot occur. By contraposition, this is equivalent to the assertion that the log scaling of $\left<b_{+}^{m}\right>^{n}$ implies $\textrm{ADV}\left(\left<b_{+}^{m}\right>^{n}\right) = 0$. Towards this, we first consider mean velocity in the inner region (including the ISL) of a general TBL flow; this obeys the law-of-the-wall scaling $\left<u_{+}\right> = f\left(y_{+}\right)$ which takes the form of log scaling in the ISL. Clearly, $\partial\left<u_{+}\right>/\partial x = \left[y\left(\textrm{d}u_{\tau}/\textrm{d}x\right)/\nu\right]\textrm{d}f/\textrm{d}y_{+}$ and $\partial\left<u_{+}\right>/\partial y = \left(u_{\tau}/\nu\right)\textrm{d}f/\textrm{d}y_{+}$. Also, putting $\left<u\right> = u_{\tau}f$ and $\partial\left<u\right>/\partial x = \left(\textrm{d}u_{\tau}/\textrm{d}x\right)\textrm{d}\left(y_{+}f\right)/\textrm{d}y_{+}$ in the mean mass conservation equation yields $\left<v\right> = -y\left(\textrm{d}u_{\tau}/\textrm{d}x\right)f$. This results in 
$\textrm{ADV}\left(\left< u_{+}\right>\right) = \left<u\right>\partial\left<u_{+}\right>/\partial x + \left<v\right>\partial\left<u_{+}\right>/\partial y = 0$ in the entire inner region irrespective of the Reynolds number $\delta_{+}$; this is true for any quantity which is a function of $y_{+}$ alone. Slope of a mean streamline in the inner region is $\textrm{d}y/\textrm{d}x = \left<v\right>/\left<u\right> = -y\left(\textrm{d}u_{\tau}/\textrm{d}x\right)/u_{\tau}$. For an iso-$y_{+}$ line, $\textrm{d}\left(yu_{\tau}\right) = 0$ i.e. $u_{\tau}\textrm{d}y + y\textrm{d}u_{\tau} = 0$. Therefore slope of an iso-$y_{+}$ line is also $\textrm{d}y/\textrm{d}x = -y\left(\textrm{d}u_{\tau}/\textrm{d}x\right)/u_{\tau}$. Physically, the equality of these slopes indicates that mean streamlines and iso-$y_{+}$ lines coincide implying no \emph{mean} exchange (`entrainment') of fluid between the outer wake region and the inner region (including ISL). Next, we consider scaling of higher-order moments in a general TBL flow. It is known that near-wall velocity fluctuations are influenced by outer-scaled motions due to the so-called amplitude modulation effects \cite{Marusicetal2010} i.e. law-of-the-wall scaling is, in general, not a correct description for variance and other higher-order moments in the inner region. Streamwise velocity fluctuation in this region can be expressed as $u'_{+} = g_{1}\left(y_{+}\right)g_{2}\left(\eta\right) + g_{3}\left(\eta\right)$ where $g_{1}$ is the universal near-wall component and $g_{2}$ and $g_{3}$ are respectively the amplitude modulation and linear superposition effects of outer-scaled motions \cite{Marusicetal2010}. This leads to a functional form $\left<b_{+}^{m}\right>^{n} = \left<u'^{m}_{+}\right>^{n} = h\left(y_{+},\eta\right)$ for a generalized higher-order moment in the inner region wherein the variables cannot be separated. However, if $\textrm{d}\left(\delta_{+}\right)/\textrm{d}x \rightarrow 0$ for the TBL flow, then slopes of iso-$y_{+}$ and iso-$\eta$ lines become equal and these lines coincide i.e. $\eta$ and $y_{+}$ no longer remain independent variables. With this, $\left<b_{+}^{m}\right>^{n} = h\left(y_{+}\right)$ - the law-of-the-wall for higher-order moments, of which log scaling is a special functional form. Clearly, as shown earlier, $\textrm{ADV}\left(\left<b_{+}^{m}\right>^{n}\right) = 0$, making this a necessary (and sufficient, as shown already) condition for comprehensive log scaling. Amplitude modulation effect in TBLs is negligible in the region where superstructures themselves reside \cite{Mathisetal2009,Mathisetal2011} and hence $y$ emerges as the only relevant lengthscale in this region (i.e. an ISL). Note that the condition $\textrm{d}\left(\delta_{+}\right)/\textrm{d}x \rightarrow 0$ is well-approximated only at ultra-high values of $\delta_{+}$ in canonical constant-pressure TBLs. This explains why in such flows, the log scaling for intensity and other higher-order moments becomes evident only at ultra-high Reynolds numbers \cite{Marusicetal2013,Vallikivietal2015} although log scaling for mean velocity has traditionally been observed all the way from moderate Reynolds numbers. In sink flows, the condition $\textrm{d}\left(\delta_{+}\right)/\textrm{d}x = 0$ is readily satisfied and comprehensive log scaling becomes apparent even at moderate Reynolds numbers.

In summary, the experimental data from sink flows demonstrate that ultra-high Reynolds number is not necessary for the occurrence of comprehensive log scaling in TBLs. This finding can be better understood in terms of the condition of streamwise invariance (zero mean-flow advection) of scaled mean velocity and higher-order moments of streamwise velocity fluctuation in the ISL. A heuristic theoretical argument shows that this condition is a necessary and sufficient condition for comprehensive log scaling in all TBL flows. 

We thank Professor R.~Narasimha for discussions. Support of the Director, Indian Institute of Tropical Meteorology (IITM), Pune is gratefully acknowledged.

\bibliographystyle{apsrev4-1}

\begin{thebibliography}{25}%
\makeatletter
\providecommand \@ifxundefined [1]{%
 \@ifx{#1\undefined}
}%
\providecommand \@ifnum [1]{%
 \ifnum #1\expandafter \@firstoftwo
 \else \expandafter \@secondoftwo
 \fi
}%
\providecommand \@ifx [1]{%
 \ifx #1\expandafter \@firstoftwo
 \else \expandafter \@secondoftwo
 \fi
}%
\providecommand \natexlab [1]{#1}%
\providecommand \enquote  [1]{``#1''}%
\providecommand \bibnamefont  [1]{#1}%
\providecommand \bibfnamefont [1]{#1}%
\providecommand \citenamefont [1]{#1}%
\providecommand \href@noop [0]{\@secondoftwo}%
\providecommand \href [0]{\begingroup \@sanitize@url \@href}%
\providecommand \@href[1]{\@@startlink{#1}\@@href}%
\providecommand \@@href[1]{\endgroup#1\@@endlink}%
\providecommand \@sanitize@url [0]{\catcode `\\12\catcode `\$12\catcode
  `\&12\catcode `\#12\catcode `\^12\catcode `\_12\catcode `\%12\relax}%
\providecommand \@@startlink[1]{}%
\providecommand \@@endlink[0]{}%
\providecommand \url  [0]{\begingroup\@sanitize@url \@url }%
\providecommand \@url [1]{\endgroup\@href {#1}{\urlprefix }}%
\providecommand \urlprefix  [0]{URL }%
\providecommand \Eprint [0]{\href }%
\providecommand \doibase [0]{http://dx.doi.org/}%
\providecommand \selectlanguage [0]{\@gobble}%
\providecommand \bibinfo  [0]{\@secondoftwo}%
\providecommand \bibfield  [0]{\@secondoftwo}%
\providecommand \translation [1]{[#1]}%
\providecommand \BibitemOpen [0]{}%
\providecommand \bibitemStop [0]{}%
\providecommand \bibitemNoStop [0]{.\EOS\space}%
\providecommand \EOS [0]{\spacefactor3000\relax}%
\providecommand \BibitemShut  [1]{\csname bibitem#1\endcsname}%
\let\auto@bib@innerbib\@empty
\bibitem [{\citenamefont {Millikan}(1938)}]{Millikan}%
  \BibitemOpen
  \bibfield  {author} {\bibinfo {author} {\bibfnamefont {C.~B.}\ \bibnamefont
  {Millikan}},\ }in\ \href@noop {} {\emph {\bibinfo {booktitle} {Proc. 5th
  Intl. Cong. Appl. Mech.}}},\ \bibinfo {editor} {edited by\ \bibinfo {editor}
  {\bibfnamefont {J.~P.}\ \bibnamefont {den Hartog}}\ and\ \bibinfo {editor}
  {\bibfnamefont {H.}~\bibnamefont {Peters}}}\ (\bibinfo  {publisher}
  {Wiley/Chapman \& Hall},\ \bibinfo {year} {1938})\ pp.\ \bibinfo {pages}
  {386--392}\BibitemShut {NoStop}%
\bibitem [{\citenamefont {Barenblatt}(1993)}]{Barenblatt1993}%
  \BibitemOpen
  \bibfield  {author} {\bibinfo {author} {\bibfnamefont {G.~I.}\ \bibnamefont
  {Barenblatt}},\ }\href@noop {} {\bibfield  {journal} {\bibinfo  {journal}
  {J.~Fluid Mech.}\ }\textbf {\bibinfo {volume} {248}},\ \bibinfo {pages} {513}
  (\bibinfo {year} {1993})}\BibitemShut {NoStop}%
\bibitem [{\citenamefont {Hultmark}\ \emph {et~al.}(2012)\citenamefont
  {Hultmark}, \citenamefont {Vallikivi}, \citenamefont {Bailey},\ and\
  \citenamefont {Smits}}]{Hultmarketal2012}%
  \BibitemOpen
  \bibfield  {author} {\bibinfo {author} {\bibfnamefont {M.}~\bibnamefont
  {Hultmark}}, \bibinfo {author} {\bibfnamefont {M.}~\bibnamefont {Vallikivi}},
  \bibinfo {author} {\bibfnamefont {S.~C.~C.}\ \bibnamefont {Bailey}}, \ and\
  \bibinfo {author} {\bibfnamefont {A.~J.}\ \bibnamefont {Smits}},\ }\href@noop
  {} {\bibfield  {journal} {\bibinfo  {journal} {Phys.~Rev.~Lett.}\ }\textbf
  {\bibinfo {volume} {108}},\ \bibinfo {pages} {1} (\bibinfo {year}
  {2012})}\BibitemShut {NoStop}%
\bibitem [{\citenamefont {Marusic}\ \emph {et~al.}(2013)\citenamefont
  {Marusic}, \citenamefont {Monty}, \citenamefont {Hultmark},\ and\
  \citenamefont {Smits}}]{Marusicetal2013}%
  \BibitemOpen
  \bibfield  {author} {\bibinfo {author} {\bibfnamefont {I.}~\bibnamefont
  {Marusic}}, \bibinfo {author} {\bibfnamefont {J.~P.}\ \bibnamefont {Monty}},
  \bibinfo {author} {\bibfnamefont {M.}~\bibnamefont {Hultmark}}, \ and\
  \bibinfo {author} {\bibfnamefont {A.~J.}\ \bibnamefont {Smits}},\ }\href@noop
  {} {\bibfield  {journal} {\bibinfo  {journal} {J.~Fluid Mech.}\ }\textbf
  {\bibinfo {volume} {716}},\ \bibinfo {pages} {R3} (\bibinfo {year}
  {2013})}\BibitemShut {NoStop}%
\bibitem [{\citenamefont {Meneveau}\ and\ \citenamefont
  {Marusic}(2013)}]{MeneveauMarusic2013}%
  \BibitemOpen
  \bibfield  {author} {\bibinfo {author} {\bibfnamefont {C.}~\bibnamefont
  {Meneveau}}\ and\ \bibinfo {author} {\bibfnamefont {I.}~\bibnamefont
  {Marusic}},\ }\href@noop {} {\bibfield  {journal} {\bibinfo  {journal}
  {J.~Fluid Mech.}\ }\textbf {\bibinfo {volume} {719}},\ \bibinfo {pages} {R1}
  (\bibinfo {year} {2013})}\BibitemShut {NoStop}%
\bibitem [{\citenamefont {Vallikivi}\ \emph {et~al.}(2015)\citenamefont
  {Vallikivi}, \citenamefont {Hultmark},\ and\ \citenamefont
  {Smits}}]{Vallikivietal2015}%
  \BibitemOpen
  \bibfield  {author} {\bibinfo {author} {\bibfnamefont {M.}~\bibnamefont
  {Vallikivi}}, \bibinfo {author} {\bibfnamefont {M.}~\bibnamefont {Hultmark}},
  \ and\ \bibinfo {author} {\bibfnamefont {A.~J.}\ \bibnamefont {Smits}},\
  }\href@noop {} {\bibfield  {journal} {\bibinfo  {journal} {J.~Fluid Mech.}\
  }\textbf {\bibinfo {volume} {779}},\ \bibinfo {pages} {371} (\bibinfo {year}
  {2015})}\BibitemShut {NoStop}%
\bibitem [{\citenamefont {Narasimha}(1990)}]{RN1990}%
  \BibitemOpen
  \bibfield  {author} {\bibinfo {author} {\bibfnamefont {R.}~\bibnamefont
  {Narasimha}},\ }in\ \href@noop {} {\emph {\bibinfo {booktitle} {Whither
  Turbulence? Turbulence at the crossroads}}},\ \bibinfo {editor} {edited by\
  \bibinfo {editor} {\bibfnamefont {J.~L.}\ \bibnamefont {Lumley}}}\ (\bibinfo
  {publisher} {Springer-Verlag},\ \bibinfo {year} {1990})\BibitemShut {NoStop}%
\bibitem [{\citenamefont {Tennekes}\ and\ \citenamefont
  {Lumley}(1972)}]{TennekesLumley}%
  \BibitemOpen
  \bibfield  {author} {\bibinfo {author} {\bibfnamefont {H.}~\bibnamefont
  {Tennekes}}\ and\ \bibinfo {author} {\bibfnamefont {J.~L.}\ \bibnamefont
  {Lumley}},\ }\href@noop {} {\emph {\bibinfo {title} {A First Course in
  Turbulence}}}\ (\bibinfo  {publisher} {MIT Press, Cambridge, MA},\ \bibinfo
  {year} {1972})\BibitemShut {NoStop}%
\bibitem [{\citenamefont {Afzal}(1976)}]{Afzal1976}%
  \BibitemOpen
  \bibfield  {author} {\bibinfo {author} {\bibfnamefont {N.}~\bibnamefont
  {Afzal}},\ }\href@noop {} {\bibfield  {journal} {\bibinfo  {journal}
  {Phys.~Fluids}\ }\textbf {\bibinfo {volume} {19}},\ \bibinfo {pages} {600}
  (\bibinfo {year} {1976})}\BibitemShut {NoStop}%
\bibitem [{\citenamefont {Townsend}(1976)}]{Townsend1976}%
  \BibitemOpen
  \bibfield  {author} {\bibinfo {author} {\bibfnamefont {A.~A.}\ \bibnamefont
  {Townsend}},\ }\href@noop {} {\emph {\bibinfo {title} {The structure of
  turbulent shear flow}}},\ \bibinfo {edition} {2nd}\ ed.\ (\bibinfo
  {publisher} {Cambridge University Press},\ \bibinfo {year}
  {1976})\BibitemShut {NoStop}%
\bibitem [{\citenamefont {Dixit}\ and\ \citenamefont
  {Ramesh}(2008)}]{DixitRamesh2008}%
  \BibitemOpen
  \bibfield  {author} {\bibinfo {author} {\bibfnamefont {S.~A.}\ \bibnamefont
  {Dixit}}\ and\ \bibinfo {author} {\bibfnamefont {O.~N.}\ \bibnamefont
  {Ramesh}},\ }\href@noop {} {\bibfield  {journal} {\bibinfo  {journal}
  {J.~Fluid Mech.}\ }\textbf {\bibinfo {volume} {615}},\ \bibinfo {pages} {445}
  (\bibinfo {year} {2008})}\BibitemShut {NoStop}%
\bibitem [{\citenamefont {Perry}\ \emph {et~al.}(2002)\citenamefont {Perry},
  \citenamefont {Marusic},\ and\ \citenamefont {Jones}}]{Perryetal2002}%
  \BibitemOpen
  \bibfield  {author} {\bibinfo {author} {\bibfnamefont {A.~E.}\ \bibnamefont
  {Perry}}, \bibinfo {author} {\bibfnamefont {I.}~\bibnamefont {Marusic}}, \
  and\ \bibinfo {author} {\bibfnamefont {M.~B.}\ \bibnamefont {Jones}},\
  }\href@noop {} {\bibfield  {journal} {\bibinfo  {journal} {J.~Fluid Mech.}\
  }\textbf {\bibinfo {volume} {461}},\ \bibinfo {pages} {61} (\bibinfo {year}
  {2002})}\BibitemShut {NoStop}%
\bibitem [{\citenamefont {Perry}\ and\ \citenamefont
  {Marusic}(1995)}]{PerryMarusic1995}%
  \BibitemOpen
  \bibfield  {author} {\bibinfo {author} {\bibfnamefont {A.~E.}\ \bibnamefont
  {Perry}}\ and\ \bibinfo {author} {\bibfnamefont {I.}~\bibnamefont
  {Marusic}},\ }\href@noop {} {\bibfield  {journal} {\bibinfo  {journal}
  {J.~Fluid Mech.}\ }\textbf {\bibinfo {volume} {298}},\ \bibinfo {pages} {361}
  (\bibinfo {year} {1995})}\BibitemShut {NoStop}%
\bibitem [{\citenamefont {Marusic}\ and\ \citenamefont
  {Perry}(1995)}]{MarusicPerry1995}%
  \BibitemOpen
  \bibfield  {author} {\bibinfo {author} {\bibfnamefont {I.}~\bibnamefont
  {Marusic}}\ and\ \bibinfo {author} {\bibfnamefont {A.~E.}\ \bibnamefont
  {Perry}},\ }\href@noop {} {\bibfield  {journal} {\bibinfo  {journal}
  {J.~Fluid Mech.}\ }\textbf {\bibinfo {volume} {298}},\ \bibinfo {pages} {389}
  (\bibinfo {year} {1995})}\BibitemShut {NoStop}%
\bibitem [{\citenamefont {Dixit}\ and\ \citenamefont
  {Ramesh}(2010)}]{DixitRamesh2010}%
  \BibitemOpen
  \bibfield  {author} {\bibinfo {author} {\bibfnamefont {S.~A.}\ \bibnamefont
  {Dixit}}\ and\ \bibinfo {author} {\bibfnamefont {O.~N.}\ \bibnamefont
  {Ramesh}},\ }\href@noop {} {\bibfield  {journal} {\bibinfo  {journal}
  {J.~Fluid Mech.}\ }\textbf {\bibinfo {volume} {649}},\ \bibinfo {pages} {233}
  (\bibinfo {year} {2010})}\BibitemShut {NoStop}%
\bibitem [{\citenamefont {Dixit}\ and\ \citenamefont
  {Ramesh}(2013)}]{DixitRamesh2013}%
  \BibitemOpen
  \bibfield  {author} {\bibinfo {author} {\bibfnamefont {S.~A.}\ \bibnamefont
  {Dixit}}\ and\ \bibinfo {author} {\bibfnamefont {O.~N.}\ \bibnamefont
  {Ramesh}},\ }\href@noop {} {\bibfield  {journal} {\bibinfo  {journal}
  {J.~Fluid Mech.}\ }\textbf {\bibinfo {volume} {737}},\ \bibinfo {pages} {329}
  (\bibinfo {year} {2013})}\BibitemShut {NoStop}%
\bibitem [{\citenamefont {McComb}(1991)}]{McComb1991}%
  \BibitemOpen
  \bibfield  {author} {\bibinfo {author} {\bibfnamefont {W.~D.}\ \bibnamefont
  {McComb}},\ }\href@noop {} {\emph {\bibinfo {title} {The Physics of Fluid
  turbulence}}},\ \bibinfo {edition} {1st}\ ed.\ (\bibinfo  {publisher}
  {Clarendon Press, Oxford},\ \bibinfo {year} {1991})\BibitemShut {NoStop}%
\bibitem [{\citenamefont {Smits}\ \emph {et~al.}(2011)\citenamefont {Smits},
  \citenamefont {Monty}, \citenamefont {Hultmark}, \citenamefont {Bailey},
  \citenamefont {Hutchins},\ and\ \citenamefont {Marusic}}]{Smitsetal2011}%
  \BibitemOpen
  \bibfield  {author} {\bibinfo {author} {\bibfnamefont {A.~J.}\ \bibnamefont
  {Smits}}, \bibinfo {author} {\bibfnamefont {J.}~\bibnamefont {Monty}},
  \bibinfo {author} {\bibfnamefont {M.}~\bibnamefont {Hultmark}}, \bibinfo
  {author} {\bibfnamefont {S.~C.~C.}\ \bibnamefont {Bailey}}, \bibinfo {author}
  {\bibfnamefont {N.}~\bibnamefont {Hutchins}}, \ and\ \bibinfo {author}
  {\bibfnamefont {I.}~\bibnamefont {Marusic}},\ }\href@noop {} {\bibfield
  {journal} {\bibinfo  {journal} {J.~Fluid Mech.}\ }\textbf {\bibinfo {volume}
  {676}},\ \bibinfo {pages} {41} (\bibinfo {year} {2011})}\BibitemShut
  {NoStop}%
\bibitem [{\citenamefont {Dengel}\ and\ \citenamefont
  {Fernholz}(1990)}]{DengelFernholz1990}%
  \BibitemOpen
  \bibfield  {author} {\bibinfo {author} {\bibfnamefont {P.}~\bibnamefont
  {Dengel}}\ and\ \bibinfo {author} {\bibfnamefont {H.~H.}\ \bibnamefont
  {Fernholz}},\ }\href@noop {} {\bibfield  {journal} {\bibinfo  {journal}
  {J.~Fluid Mech.}\ }\textbf {\bibinfo {volume} {212}},\ \bibinfo {pages} {615}
  (\bibinfo {year} {1990})}\BibitemShut {NoStop}%
\bibitem [{\citenamefont {Fernholz}\ and\ \citenamefont
  {Finley}(1996)}]{FernholzFinley}%
  \BibitemOpen
  \bibfield  {author} {\bibinfo {author} {\bibfnamefont {H.~H.}\ \bibnamefont
  {Fernholz}}\ and\ \bibinfo {author} {\bibfnamefont {P.~J.}\ \bibnamefont
  {Finley}},\ }\href@noop {} {\bibfield  {journal} {\bibinfo  {journal}
  {Prog.~Aerospace Sci.}\ }\textbf {\bibinfo {volume} {32}},\ \bibinfo {pages}
  {245} (\bibinfo {year} {1996})}\BibitemShut {NoStop}%
\bibitem [{\citenamefont {Huisman}\ \emph {et~al.}(2013)\citenamefont
  {Huisman}, \citenamefont {Scharnowski}, \citenamefont {Cierpka},
  \citenamefont {K\mbox{\"{a}}hler}, \citenamefont {Lohse},\ and\ \citenamefont
  {Sun}}]{Huismanetal2013}%
  \BibitemOpen
  \bibfield  {author} {\bibinfo {author} {\bibfnamefont {S.~G.}\ \bibnamefont
  {Huisman}}, \bibinfo {author} {\bibfnamefont {S.}~\bibnamefont
  {Scharnowski}}, \bibinfo {author} {\bibfnamefont {C.}~\bibnamefont
  {Cierpka}}, \bibinfo {author} {\bibfnamefont {C.~J.}\ \bibnamefont
  {K\mbox{\"{a}}hler}}, \bibinfo {author} {\bibfnamefont {D.}~\bibnamefont
  {Lohse}}, \ and\ \bibinfo {author} {\bibfnamefont {C.}~\bibnamefont {Sun}},\
  }\href@noop {} {\bibfield  {journal} {\bibinfo  {journal} {Phys.~Rev.~Lett.}\
  }\textbf {\bibinfo {volume} {110}},\ \bibinfo {pages} {1} (\bibinfo {year}
  {2013})}\BibitemShut {NoStop}%
\bibitem [{\citenamefont {Monkewitz}\ and\ \citenamefont
  {Nagib}(2015)}]{MonkewitzNagib2015}%
  \BibitemOpen
  \bibfield  {author} {\bibinfo {author} {\bibfnamefont {P.~A.}\ \bibnamefont
  {Monkewitz}}\ and\ \bibinfo {author} {\bibfnamefont {H.~M.}\ \bibnamefont
  {Nagib}},\ }\href@noop {} {\bibfield  {journal} {\bibinfo  {journal}
  {J.~Fluid Mech.}\ }\textbf {\bibinfo {volume} {783}},\ \bibinfo {pages} {474}
  (\bibinfo {year} {2015})}\BibitemShut {NoStop}%
\bibitem [{\citenamefont {Marusic}\ \emph {et~al.}(2010)\citenamefont
  {Marusic}, \citenamefont {Mathis},\ and\ \citenamefont
  {Hutchins}}]{Marusicetal2010}%
  \BibitemOpen
  \bibfield  {author} {\bibinfo {author} {\bibfnamefont {I.}~\bibnamefont
  {Marusic}}, \bibinfo {author} {\bibfnamefont {R.}~\bibnamefont {Mathis}}, \
  and\ \bibinfo {author} {\bibfnamefont {N.}~\bibnamefont {Hutchins}},\
  }\href@noop {} {\bibfield  {journal} {\bibinfo  {journal} {Science}\ }\textbf
  {\bibinfo {volume} {329}},\ \bibinfo {pages} {193} (\bibinfo {year}
  {2010})}\BibitemShut {NoStop}%
\bibitem [{\citenamefont {Mathis}\ \emph {et~al.}(2009)\citenamefont {Mathis},
  \citenamefont {Hutchins},\ and\ \citenamefont {Marusic}}]{Mathisetal2009}%
  \BibitemOpen
  \bibfield  {author} {\bibinfo {author} {\bibfnamefont {R.}~\bibnamefont
  {Mathis}}, \bibinfo {author} {\bibfnamefont {N.}~\bibnamefont {Hutchins}}, \
  and\ \bibinfo {author} {\bibfnamefont {I.}~\bibnamefont {Marusic}},\
  }\href@noop {} {\bibfield  {journal} {\bibinfo  {journal} {J.~Fluid Mech.}\
  }\textbf {\bibinfo {volume} {628}},\ \bibinfo {pages} {311} (\bibinfo {year}
  {2009})}\BibitemShut {NoStop}%
\bibitem [{\citenamefont {Mathis}\ \emph {et~al.}(2011)\citenamefont {Mathis},
  \citenamefont {Marusic}, \citenamefont {Hutchins},\ and\ \citenamefont
  {Sreenivasan}}]{Mathisetal2011}%
  \BibitemOpen
  \bibfield  {author} {\bibinfo {author} {\bibfnamefont {R.}~\bibnamefont
  {Mathis}}, \bibinfo {author} {\bibfnamefont {I.}~\bibnamefont {Marusic}},
  \bibinfo {author} {\bibfnamefont {N.}~\bibnamefont {Hutchins}}, \ and\
  \bibinfo {author} {\bibfnamefont {K.~R.}\ \bibnamefont {Sreenivasan}},\
  }\href@noop {} {\bibfield  {journal} {\bibinfo  {journal} {Phys.~Fluids}\
  }\textbf {\bibinfo {volume} {23}},\ \bibinfo {pages} {1} (\bibinfo {year}
  {2011})}\BibitemShut {NoStop}%
\end{thebibliography}
%

\end{document}